# Aperiodic quantum oscillations in the two-dimensional electron gas at the LaAlO$_3$/SrTiO$_3$ interface

Km Rubi,* Michel Goiran, and Walter Escoffier
*LNCMI-EMFL, CNRS, INSA, UPS, 143 Avenue de Rangueil, 31400 Toulouse, France*

Julien Gosteau, Rémi Arras, Benedicte Warot-Fonrose, Raphaël Serra, and Etienne Snoeck
*CEMES, Université de Toulouse, CNRS, UPS, 29, rue Jeanne-Marvig, F-31055 Toulouse, France*

Kun Han, Shengwei Zeng, Zhen Huang, and Ariando†
*Department of Physics and NUSNNI-Nanocore, National University of Singapore, 117411 Singapore*
(Dated: February 27, 2019)

Despite several attempts, the intimate electronic structure of two-dimensional electron systems buried at the interface between LaAlO$_3$ and SrTiO$_3$ still remains to be experimentally revealed. Here, we investigate the transport properties of a high-mobility quasi-two-dimensional electron gas at this interface under high magnetic field (55 T) and provide new insights for electronic band structure by analyzing the Shubnikov-de Haas oscillations. Interestingly, the quantum oscillations are not $1/B$-periodic and produce a highly non-linear Landau plot (Landau level index versus $1/B$). Among possible scenarios, the Roth-Gao-Niu equation provides a natural explanation for $1/B$-aperiodic oscillations in relation with the magnetic response functions of the system. Overall, the magneto-transport data are discussed in light of high-resolution scanning transmission electron microscopy analysis of the interface as well as calculations from density functional theory.

**Introduction**

The discovery of a two-dimensional electron gas (2DEG) at the interface between two insulators LaAlO$_3$ (LAO) and SrTiO$_3$ (STO) [1] has not only enhanced the expectations in oxide-electronics but has also brought new and exciting opportunities to explore the novel physics of two-dimensional electron gas with unmapped parameters. In recent years, this interface has been studied extensively and several exotic phenomena including superconductivity [2–4], Rashba spin-orbit coupling [5–7], ferromagnetism [8–10], and quantum oscillations [11–14] have been reported. The leading consensus for the formation of the 2DEG at this interface is the electronic reconstruction (also refered to as the "polar catastrophe") [15, 16]. This scenario is based on electron transfer from the polar LAO to the interfacial layers of STO, although other mechanisms such as oxygen vacancy doping [17, 18] or interdiffusion [19] are also at play.

First-principles calculations of the band-structure reveal the occupancy of several non-degenerate sub-bands d$_{xy}$, d$_{xz}$ and d$_{yz}$ originated from crystal field splitted Ti:3d-t$_{2g}$ orbitals located at the interface or in its vicinity [20]. Using combination of k.p envelope function method and density functional theory, Heeringen *et al.* [21] confirmed the existence of many anisotropic and non-parabolic sub-bands, which are only a few meV apart. Meanwhile, Zhong *et al.* [6] predicted a large spin orbit splitting at the crossing point of the d$_{xy}$ and d$_{xz/yz}$ bands using density functional theory. To date, the predicted band-structure has received only partial support from transport experiments. Indeed, the analysis of Shubnikov-de Haas Oscillations (SdHO) provided distinctive results with respect to different experimental conditions and sample preparation methods. In references [11] and [12], the authors reported independently a single SdHO frequency and pointed out the possible valley and/or spin degeneracy of the electronic states to account for the discrepancy between the carrier density extracted from the SdHO and the Hall effect. Later, double SdHO frequencies perceived up to a moderate magnetic field (8 T) and at low temperature (50 mK) were attributed to carriers populating the heavy sub-bands d$_{xz}$ and d$_{yz}$ splitted by the Rashba spin-orbit interaction [22]. On the other hand, McCollam *et al.* [13] extracted four frequencies using high magnetic field (20 T) for the SrTiO$_3$/SrCuO$_3$/LaAlO$_3$/SrTiO$_3$ hetero-structure, and assigned these frequencies to four sub-bands of different mobilities. In contrast, very high field (55 T) measurements reveal quasi-$1/B$ periodic oscillations with an apparent single frequency, which emphasizes the role of Rashba spin-orbit interaction [14]. No consensus is reached in the interpretation of the magneto-transport properties and further work is needed to enrich the debate. In addition, it is worth mentioning that a systematic discrepancy between the carrier density estimated from Hall effect ($n_{Hall}$) and SdHO ($n_{SdHO}$) is reported by several groups [11, 12, 14].

Here, we investigate the transport properties of a high mobility ($\mu > 10^4$ cm$^2$V$^{-1}$s$^{-1}$) 2DEG at the LAO/STO interface under very high magnetic field (55 T) and low temperature (1.6 K). We observe strong $1/B$-aperiodic SdHO where the frequency increases with magnetic field. The origin and hypothesis leading to this unexpected effect are discussed below, after reviewing the experimental data.

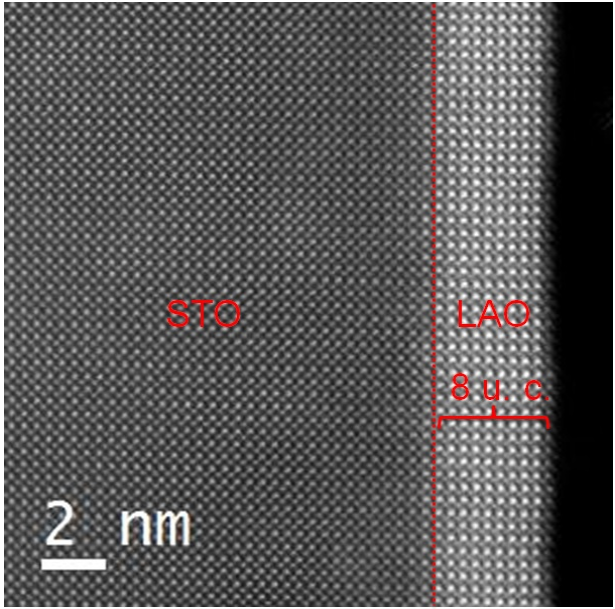

FIG. 1: High angle annular dark field image of the LAO(8 u.c.)/STO interface, using a high-resolution scanning transmission electron microscope (HRSTEM). No misfit dislocation detected at the interface (depicted by the red dashed line) indicates that the LAO layer is epitaxially grown on STO.

**Results**

The sample under investigation consists of 8 unit cells of LAO grown on top of a 500 $\mu$m STO (001) substrate using a pulsed laser deposition technique at the temperature of 760$^o$C and in $2 \times 10^{-6}$ Torr oxygen pressure. After the LAO growth, the sample is annealed in $6 \times 10^{-2}$ Torr oxygen pressure at a temperature of 600$^o$C to remove oxygen vacancies. Fig.1 shows an atomic resolution image of the interface. While the LAO layer is mono-crystalline without surface roughness, a large cation intermixing extending within two-three unit cells from the interface in both LAO and STO is observed. The lattice spacing is expanded of about 5.5% along the growth axis. However, no distortion is observed in the growth plane. The sample is tailored into a Hall bar of width 50 $\mu$m and length 160 $\mu$m (see supplementary information S1). Both the longitudinal magneto-resistance ($R_{xx}$) and the Hall resistance ($R_{xy}$) are measured simultaneously during a pulse of magnetic field up to 55 T at a temperature $T = 1.6$ K. Such measurements are repeated for several values of back-gate voltage $V_g$, which is applied at the rear face of the STO substrate to modulate the carrier density of the 2DEG. The panels (a) and (b) of Fig.2 show the magnetic field dependence of $R_{xx}$ and $R_{xy}$, respectively, recorded at different back-gate voltages from 0 to $-15.5$ V. The depopulation of the 2DEG leads to a substantial increase in zero-field resistance as well as an increasing positive magneto-resistance. In contrast, the longitudinal and Hall resistances remain unchanged as the back-gate voltage is set at positive values in the range $\sim 5$ to 60 V as shown in Fig.2-(d) and (e). After subtracting the monotonic background (see supplementary information S2), we show $\Delta R_{xx}(B) = R_{xx}(B) - \langle R_{xx}(B) \rangle$ for various $V_g$ in Fig.2-(c) and (f), respectively. While the position of the extrema in $\Delta R_{xx}(B)$ spectra shifts significantly towards a lower magnetic field for negative back-gate voltages, no deviation is noticed above $V_g \sim 5$ V, which reflects a saturation of the carrier density.

The Hall carrier concentration $n_{Hall}$ and mobility $\mu_{Hall}$ are plotted versus $V_g$ in Fig.3. Taking into account the electric field dependence of the STO dielectric constant [23], the plane capacitor model (supplementary information S3) reproduces well the experimental Hall carrier density in the back-gate voltage range $-7$ V$< V_g < 5$ V. For negative back-gate voltages, the mobility decreases due to the enhanced electron's sensitiveness to emerging local potential fluctuations. Indeed, as the carrier density becomes low, screening of charged defects in the close environment of the 2DEG vanishes, leading to a rise of the scattering rate. For $V_g < -7$ V, the Hall voltage shows large temporal fluctuations and eventually becomes unmeasurable (see dotted-lines in Fig.2-(b)), as if the sample was turn into an insulator. This reversible trend is confirmed by $R_{xx}$ which abruptly exceeds the measurement limit ($\sim 1$M$\Omega$) for $V_g < -15.5$ V. In this regime, i. e. when the Hall 2DEG carrier density is below the threshold value of $\sim 0.98 \times 10^{13}$ cm$^{-2}$, electron localization arises successively in different locations of the sample, leading to a transport regime governed by the presence of percolating conduction paths. This scenario, in agreement with the direct visualization of sub-micrometer electron puddles using scanning SQUID spectroscopy [9] and inter-modulation electrostatic force microscopy [24], provides a natural explanation for the apparent mismatch between $R_{xx}$ and $R_{xy}$ back-gating insulating thresholds. On the other hand, when the back-gate voltage is pushed beyond $+10$ V, the Hall carrier density saturates to a maximum value of $\sim 1.95 \times 10^{13}$ cm$^{-2}$. This response is accompanied by a strong hysteretic behaviour, which originates from the trapping of electrons by defects [25, 26]. Indeed, assuming that the 2DEG is confined in a potential well close to the interface with the Fermi energy lying close to the top of the well, a positive back-gate voltage modifies the well profile and bends the STO conduction band. As a consequence, electrons can irreversibly escape the well and get trapped by defects. Interestingly, the carrier density can be reset to its $V_g = 0$ V initial value (after a round trip to +60 V at 1.6 K) by warming up the sample slightly above 20 K. We thus infer a very small average trapping energy of the order of 2 meV in our sample. We conjecture that the growing number of ionized defects, acting as remote scattering centers, is responsible for the variation of the Hall mobility which decreases simultaneously with the onset of carrier density saturation. Oxygen vacancies are often cited as potential defects in oxides but, eventhough investigations on the nature of the defects is certainly beyond the scope of this study, we believe that they are not at play in the electron trapping process. Indeed, first we remind that our sample was annealed in oxygen atmosphere and the resulting carrier density is lower than that typically obtained from the LAO/STO interfaces prepared in a low oxygen partial pressure without a subsequent high pressure post-annealing. Second, the oxygen vacancies are positively charged when they transfer one or two electrons to the 2DEG. The capture of electrons with increasing back-gate voltage would render them neutral, which is inconsistent with the experimentally observed mobility decrease.

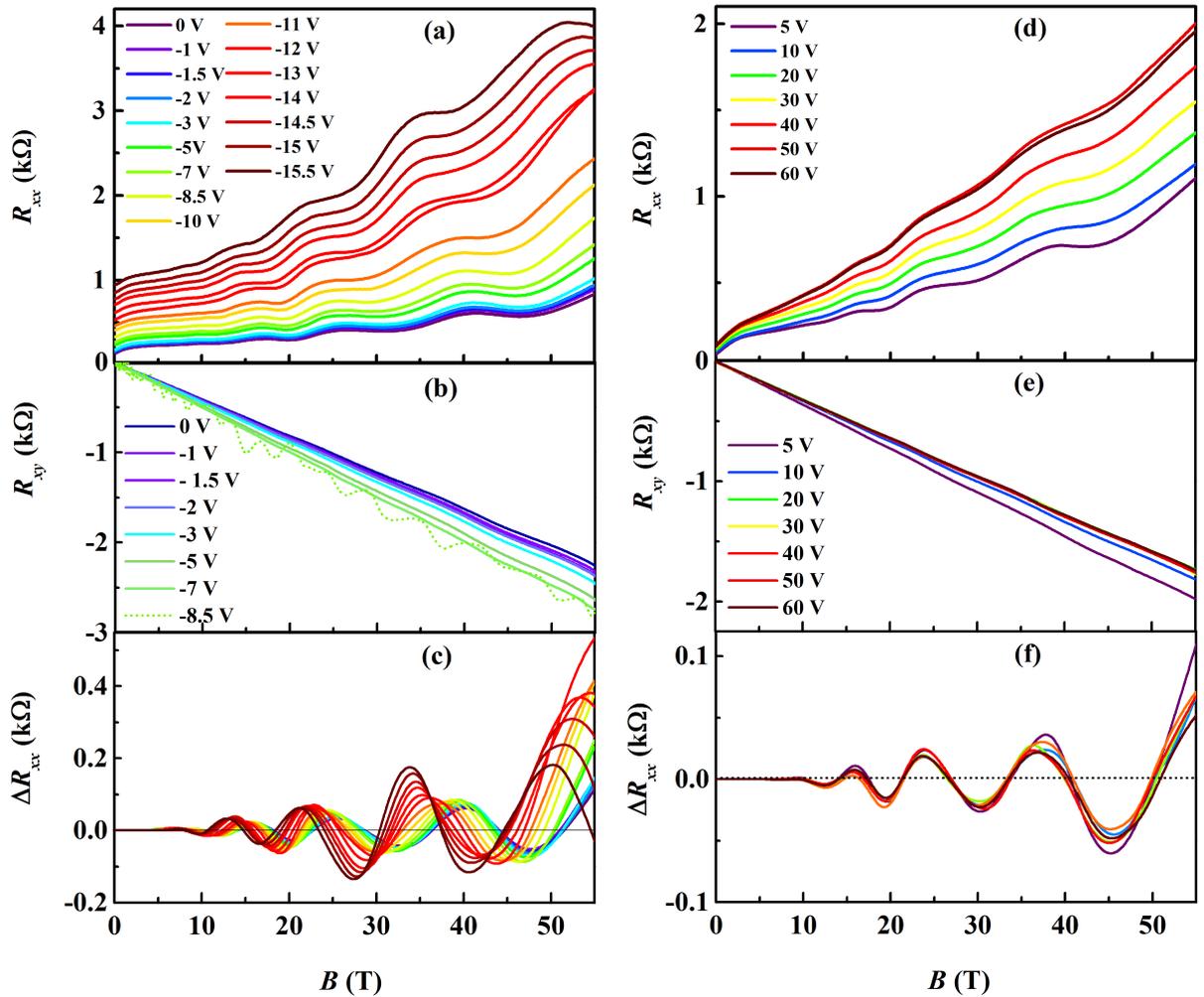

FIG. 2: Left panels: magnetic field dependence of (a) $R_{xx}$, (b) $R_{xy}$ and (c) $\Delta R_{xx}$ for different values of $V_g$ in the range $0 \to -15.5\ V$. Right panels: magnetic field dependence of (d) $R_{xx}$, (e) $R_{xy}$ and (f) $\Delta R_{xx}$ for different values of $V_g$ in the range $5 \to +60\ V$. The temperature is 1.6 K. $\Delta R_{xx}$ is the oscillating part of $R_{xx}$ after subtracting the monotonous background. A significant back-gate induced shift in the position of the $\Delta R_{xx}$ oscillations extrema is observed only for $V_g < 5\ V$. Above this value, the oscillation pattern is almost unchanged concurrent with the saturation of $n_{Hall}(V_g)$.

The main panel of Fig.4-(a) focuses on the inverse magnetic field dependence $\Delta R_{xx}$ for $V_g = 0$ V. According to the Onsager relation, the SdHO are expected to be $1/B$-periodic. However we notice that the frequency of the oscillations increases monotonically with increasing magnetic field in our LAO/STO sample. This unexpected effect becomes clearer in the Landau plot, where the extrema of the oscillations are plotted as a function of $1/B$ in the inset of Fig.4-(a). Interestingly, the Landau plot is highly nonlinear, contrary to the standard 2DEG in semiconductor hetero-structures or in graphene. As a matter of fact, the SdHO aperiodicity was apparent in the early literature of quantum transport in LAO/STO systems above 1 K [11, 27]. However, this evidence has probably been overlooked due to the narrow magnetic field range and limited number of oscillations. Similarly, $\delta$-doped STO also exhibits aperiodic oscillations up to $B_{max} = 15$ T in reference [28]. In the following, we review and comment the different scenarios leading to $1/B$-aperiodic SdHO.

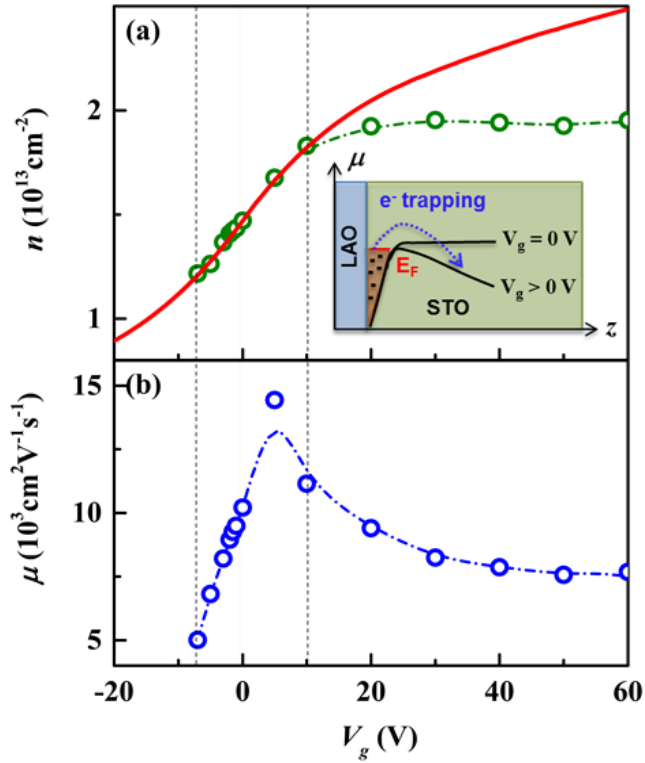

FIG. 3: (a) Charge carrier density estimated from the Hall resistance (green symbols) and the plane capacitance model (red solid line). (b) Electron mobility estimated from the Hall density and the zero-field longitudinal. The dash lines are guides for the eyes. The region between two vertical dotted lines (-7 V < $V_g$ < 10 V) indicates a reversible back-gate induced electron density modulation. The right side of the vertical line represents an hysteretic charge trapping/detrapping regime by ionized defects, leading to a saturation of the carrier density. The inset depicts an illustration of the LAO/STO interface including confining potential and electron trapping process.

**Discussion**

First, we recall that in the context of the Onsager's quantization condition, the frequency of the magnetic oscillations is directly linked to the carrier density. We therefore conjecture that electrons may be continuously injected in the 2DEG as the magnetic field increases. Similar to graphene on silicon carbide, an hypothetical reservoir of charges may exist close to the interface from which electrons are transferred to the 2DEG [29, 30]. Although the experimental data can be nicely reproduced using the Lifshitz-Kosevich equation, where a phenomenological magnetic field dependent 2D carrier density $n_{(B)}^{2D}$ is used, the model predicts a relative increase of $n_{(B)}^{2D}$ between 200% and 1000% over the full magnetic field range depending on the initial carrier concentration at $B = 0$ T (see supplementary information S4). This hypothesis is therefore discarded, since such results are improbable and inconsistent with the observed linear Hall resistance. Assuming a constant density of charges, a magnetic field-induced change of the Fermi surface topology is a conceivable alternative to explain the $1/B$-aperiodic oscillations. This scenario is supported by the presence of strong atomic and Rashba spin-orbit couplings, which introduces complex band dispersion at the Fermi energy. Beyond this qualitative approach, dedicated theoretical calculations are required to understand the resulting Landau level formation under high magnetic field. Another scenario involves the onset of spin degeneracy lifting. As pointed out by F. Trier *et al.* [27], the orbital Landau level gap $\hbar eB/m^*$ is comparable to the Zeeman energy $g\mu_B.B$ for large effective mass ($m^* \sim 1.3 \times m_e$). This model predicts an unusual magnetic field doubling of the oscillation's frequency corresponding to the transition from a spin degenerate to a spin-resolved Landau level ladder. Although this hypothesis sounds attractive, it fails to explain our results, since the period of the oscillations is actually divided by 3.6 from 12 to 34 T. Very recently, G. Cheng *et al.* [31] studied the conductance of an artificially designed quasi-1D conduction channel at the LAO/STO interface, where the competition between electrical and magnetic confinement yields $1/B$-aperiodic oscillations. While this model can merely fit our experimental data, such an hypothesis actually requires a incoherent network of quasi-1D conduction channels at the interface, whose origin remains elusive. Besides, this model brings about several issues regarding the fitting parameters such as the carrier density, the conduction channel width, the effective mass or the confining potential profile (see supplementary information S5).

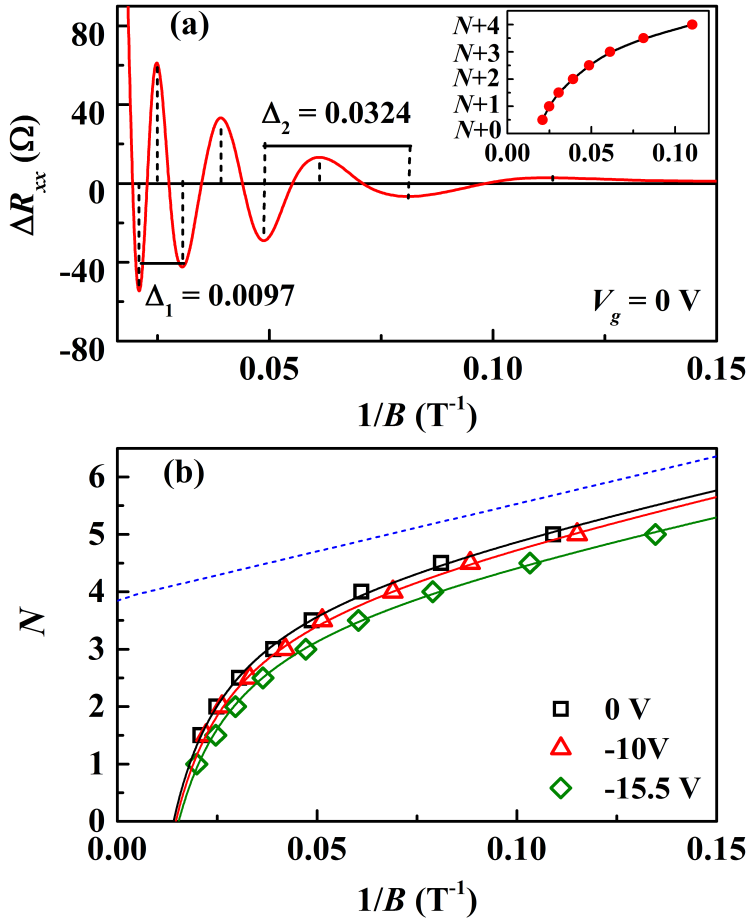

FIG. 4: (a) Main panel: Inverse field dependence of $\Delta R_{xx}$ for $V_g = 0$ V. Inset: the corresponding Landau plot. Maxima and minima of $\Delta R_{xx}(1/B)$ are assigned to integer and half-integer indices, respectively. (b) Landau level plots (symbols) for $V_g = 0$ V, -10 V and -15.5 V with Roth-Gao-Niu quantization fit (solid lines). The dash line represents the asymptote for $V_g = 0$ V derived from the fitting parameters $f(E)$ and $\gamma_0(E)$.

| References | $n_{Hall}$ (cm$^{-2}$) | $f(E)$ | $\gamma_0(E)$ | $C(E)$ | $n_{SdHO}$ (cm$^{-2}$) |
|---|---|---|---|---|---|
| This work ($V_g$=0V) | $1.47 \times 10^{13}$ | 16.56 | -3.87 | -0.057 | $8.02 \times 10^{11}$ |
| This work ($V_g$=-10 V) | $1.12 \times 10^{13}$ | 15.74 | -3.76 | -0.058 | $7.62 \times 10^{11}$ |
| This work ($V_g$=-15.5 V) | $0.98 \times 10^{13}$ | 11.68 | -3.61 | -0.062 | $5.66 \times 10^{11}$ |
| [11] | $1.05 \times 10^{13}$ | 14.49 | -3.62 | -0.19 | $6.96 \times 10^{11}$ |
| [22] | $0.56 \times 10^{13}$ | 8.94 | -3.68 | -0.24 | $4.28 \times 10^{11}$ |

TABLE I: Hall carrier density ($n_{Hall}$), fitting parameters ($f(E)$, $\gamma_0(E)$ and $C(E)$) from the fit of the experimental data (this work - Fig.3-(b) and literature - supplementary information S6) to the Eq. (2) and SdHO carrier density ($n_{SdHO}$). $n_{SdHO}$ is estimated using Onsager's relation $n_{SdHO} = \frac{2 \cdot e}{h} f_{SdHO}$. $n_{Hall}$ for $V_g$= -10 V and -15.5 V is taken from the extrapolated plane capacitor model fitting in Fig.3-(a).

Recently, Gao and Niu [32] revisited the theoretical work of Roth [33] for quantization of the energy levels of the Bloch bands and introduced a generalized Onsager's relation to explain the $1/B$-aperiodic quantum oscillations observed in systems with complex topological order. J.-N. Fuchs et al. [34] adapted the same theory and investigated the two-dimensional Dirac systems having Zeeman effect or Rashba spin-orbit interaction thoroughly. The generalized Onsager's relation, also called Roth-Gao-Niu quantization relation, is written as

$$S(E) = \frac{(2\pi)^2 \cdot e \cdot B}{h}[N + \gamma(E, B)] \qquad (1)$$

where $N$ is the Landau level index, $S(E)$ is the area in reciprocal space of the cyclotron orbit at energy $E$ and $\gamma(E, B)$

is an energy and magnetic field dependent phase. After power series expansion of $\gamma(E, B)$, this relation is written in the form :

$$N = \frac{f(E)}{B} - \gamma_0(E) + C(E) \times B + ... \qquad (2)$$

Here, $f(E)$ is the frequency of the oscillations, $\gamma_0(E) = \frac{1}{2} - \frac{\partial M}{\partial \mu}$ is the derivative of the magnetization with respect to the chemical potential the coefficient $C(E) = \frac{\phi_0}{2} \frac{\partial \chi}{\partial \mu}$ is linked to the derivative of the magnetic susceptibility with respect to the chemical potential. In perfectly parabolic or linear dispersion relations, $\frac{\partial \chi}{\partial \mu}$ turn out to be exactly zero, leading to linear Landau plot. However, a non-linear Landau plot is expected in systems involving a complex dispersion relation at the Fermi energy. First-principles calculations for LAO/STO interface have indeed predicted a strongly distorted band-structure, due to atomic and Rashba spin-orbit couplings. The Roth-Gao-Niu relation provides thus a natural explanation for the $1/B$-aperiodic oscillations at the LAO/STO interface. The fit of the experimental Landau plot is shown in Fig.4-(b) and the fitting parameters are tabulated in Table I.

The frequency of the oscillations, $f(E)$, is the slope of the Landau diagram asymptote at low field. The phase $\gamma_0(E)$ corresponds to the intercept of the asymptote with the Landau level index axis and $C(E)$ is linked to the curvature of the Landau plot, regardless the arbitrary absolute Landau level index. Interestingly, all the parameters exhibit the same gate voltage dependence (see supplementary information S7). While $f(E)$ and $\gamma_0(E)$ from high (55 T - this work) and low field (12 T - [11, 22]) experimental data are comparable, a small discrepancy occurs in $C(E)$ for different field regimes. The deviation in $C(E)$ can be explained by (i) different levels of disorder, (ii) different carrier density (different position of the Fermi energy in the band structure), and (iii) magnetic field range over which the Roth-Gao-Niu fit is performed. It is worth mentioning that the accuracy of $f(E)$ and $\gamma_0(E)$ is actually restricted by the limited number of oscillations at low field. Even if the recent observation of a non-zero magnetic response [3, 9, 35] of the LAO/STO interface is in line with the non-linear Landau plots, improved precision on $\gamma_0(E)$ and $C(E)$ is needed in order to obtain a quantitative agreement.

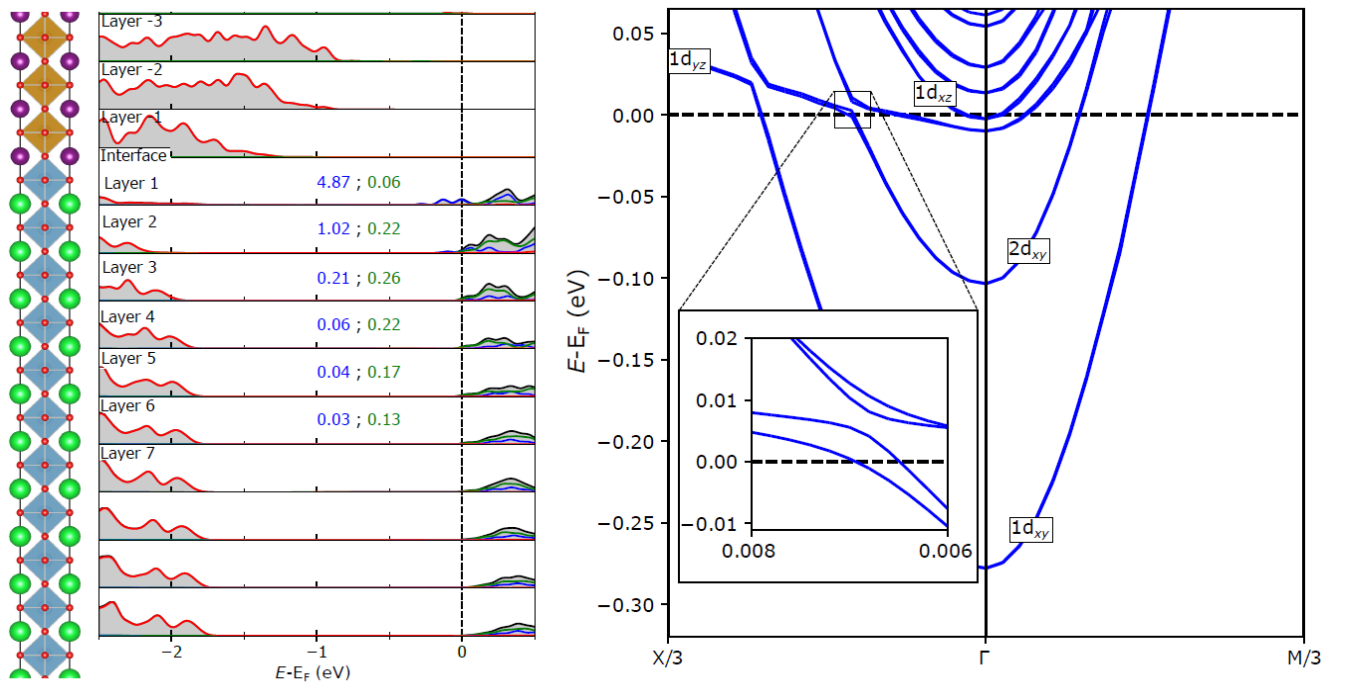

FIG. 5: Left panel: The layer-resolved density of states (LDOS) in LAO/STO. Each LDOS corresponds to a AlO$_2$ or TiO$_2$ plane. The black curve corresponds to the total LDOS of the TiO$_2$ plane, the blue, green and red curves correspond respectively to the $d_{xy}$ orbital, the $d_{yz} + d_{xz}$ orbitals of Ti and the 2p states of oxygen. Right panel: Electronic band structure of the STO in LAO/STO system. Inset shows a zoomed-in view of band avoiding and splitting due to the atomic and Rashba spin-orbit coupling.

The Roth-Gao-Niu model provides a new approach to compute the carrier density from aperiodic oscillations which was otherwise systematically over-estimated using fast Fourier transform (FFT) analysis. Indeed, the strong aperiodicity tends to broaden the FFT spectrum towards high frequency and can even favor the onset of spurious FFT peaks when the magnetic field range is extended towards very high magnetic field (see supplementary information S8). To complement this insight, we revisit the large discrepancy between $n_{Hall}$ and $n_{SdHO}$ using density functional theory (DFT) calculations, where the density of states projection of the $d_{xy}$ and $d_{xz}/d_{yz}$ subbands at the Fermi energy is investigated. The layer resolved density of states with electronic band structure of the LAO/STO interface are shown in Fig. 5. The lower lying $d_{xy}$ subbands ($1d_{xy}$ and $2d_{xy}$) are populated by a majority of electrons located mainly into the first two TiO$_2$ planes from the interface. On the other hand, $d_{xz}/d_{yz}$ subbands are populated by a minority of electrons whose spatial extension expands from the $2^{nd}$ to the $6^{th}$ TiO$_2$ plane. Based on TEM analysis, we conjecture that cation intermixing, strain and probably the non-homogeneous valence of the titanium atoms generate potential fluctuations in the two first layers of STO at least, which strongly affect the electronic mobility. Therefore, the majority of electrons belonging to the $1d_{xy}$ and $2d_{xy}$ subbands will have low mobility even if they display low effective mass and do not engender quantum oscillations. They contribute mainly to the linear Hall effect (see supplementary information S9) with a theoretical carrier density estimated to $n_{d_{xy}} \approx 6 \times 10^{13}$ cm$^{-2}$. The SDHO ensue from the minority charge carriers located deeper in STO and belonging to the higher energy $d_{xz}/d_{yz}$ subbands. The cyclotron mass, $m_c = 1.75 + 0.05 m_e$ estimated from the Lifshitz-Kosevich equation fitting (see supplementary information S10) is consistent with the theoretically calculated averaged effective mass ($m^* \sim 1.8 m_e$) of the $d_{xz}/d_{yz}$ subbands. The theoretical carrier density for these subbands is estimated to $n_{d_{xz/yz}} \approx 5.2 \times 10^{12}$ cm$^{-2}$ taking into account $4^{th}$, $5^{th}$ and $6^{th}$ layer. Even if the DFT based model does not reflect a realistic LAO/STO system (for instance the presence of defects is not into account and the exact position of the Fermi energy is unkown), this study shows that a large ratio of $n_{Hall}$ and $n_{SdHO}$ ($n_{Hall}/n_{SdHO} \sim 13 - 17$) observed experimentally can be conveniently explained by the relative population of the $d_{xy}$, $d_{xz}$ and $d_{yz}$ subbands, where $n_{total}/n_{d_{xz/yz}} \sim 14$.

**Summary**

The magneto-transport properties of a high-mobility electron gas at the LAO/STO interface are studied under high magnetic field (55 T) and low temperature (1.6 K) for different carrier densities. We observe $1/B$-aperiodic SdHO over the full magnetic field range (9 T - 55 T), which can not be explained by a magnetic field dependent carrier density. The aperiodic character of the oscillations involves the failure of standard FFT data processing, unless the analysis is restricted to low enough magnetic field. Among several possible scenarios, the Roth-Gao-Niu model provides new insights regarding the electronic properties and brings evidence for non-trivial dispersion relations at the Fermi energy. This hypothesis is consistent with DFT calculations, where $d_{xy}$, $d_{xz}$ and $d_{yz}$ subbands avoid each other due to spin-orbit coupling. When the Fermi energy is located in the vicinity of the $d_{xz}/d_{yz}$ subbands bottom, a fair correspondence can be established between electrons' location within the TiO$_2$ planes and the subbands to whom they belong. The majority electrons populating the $d_{xy}$ subbands are located close to the interface and are sensitive to interface disorder as revealed by HRSTEM analysis. Their low mobility and high carrier density explain the linear field dependence of the Hall resistance. On the other hand, the minority of charge carriers belonging to the d$_{xz}$/d$_{yz}$ bands and located deeper from the interface are responsible for the SdHO with strong aperiodic behavior. In this scheme, the long-standing issue regarding the mismatch in carrier density extracted from the Hall effect and the SdHO finds a qualitative explanation, based on the sample's dependent relative population of the d$_{xz}$, d$_{yz}$ and d$_{xy}$ subbands.

This study has been partially supported through the grant NEXT n$^o$ANR-10-LABX-0037 in the framework of the "Programme des Investissements d'Avenir". The high magnetic field and low temperature measurements were performed at LNCMI-Toulouse under EMFL proposal TSC02-217. The NUS authors acknowledge the support from the NUS Academic Research Fund (AcRF Tier 1 Grants No. R-144-000-364-112, No. R-144-000-391-114, and No. R-144-000-403-114) and the Singapore National Research Foundation (NRF) under the Competitive Research Programs (CRP Grant No. NRF-CRP15-2015-01). This work was granted access to the HPC resources of the CALMIP supercomputing center under the allocation p1229. We are grateful to U. Zeitler, M.O. Goerbig, J. N. Fuchs, F. Piechon and D. K. Maude for very insightful discussions.

**Methods**

**Sample fabrication and structural characterization:** The 8 unit cells of LaAlO$_3$ film were grown on a TiO$_2$-terminated (001) SrTiO$_3$ single crystal using pulse laser deposition (PLD) technique at NUSNNI-Nanocore National University of Singapore (NUS). The Lambda Physik Excimer KrF UV laser with the wavelength of 248 nm at a pulse rate of 1 Hz was used in PLD. The layer-by-layer growth of LaAlO$_3$ was precisely controlled by in-situ reflection high energy electron diffraction (RHEED). During growth, the oxygen partial pressure $P_{(O_2)}$ and temperature were maintained at $2 \times 10^{-6}$ torr and 760$^o$C, respectively. After growth, the sample was cooled down to room temperature at a rate of 5$^o$C/min with same $P_{(O_2)}$. Later, the sample was annealed in the $P_{(O_2)}$ of 60 mtorr and at the temperature of 600$^o$C. The six-terminal Hall bar device of width 50 $\mu$m and length 160 $\mu$m between the longitudinal voltage leads was fabricated by conventional photolithography technique using AlN films as hard mask. The optical microscopy image of the Hall bar device is shown in the supplementary information S1. The cross-section of the LAO/STO structure was characterized using a probe Cs-corrected JEOL JEM-ARM200F Transmission Electron Microscope (TEM) operated at 200 kV. Focused ion beam was used to prepare the TEM lamella.

**Transport measurements:** The transport measurements under high magnetic field and low temperatures were carried out at Laboratoire National des Champs Magnétiques Intenses (LNCMI) Toulouse. For transport measurements, the device was electrically bonded with Al wires using wedge bonding. The sample was cooled down from room temperature to 1.6 K with a cooling rate of $\sim 3$ K/min in a He4 cryostat. The sheet resistance decreases from 250 k$\Omega$ to 130 $\Omega$ as the temperature is decreased from 300 K to 1.6 K. The simultaneous measurements of longitudinal and Hall voltages were performed under high pulsed magnetic field of up to 55 T with duration 300 ms and at $T = 1.6$ K. The back-gate voltage was varied from 0 V to $-15.5$ V and from 0 V to 60 V for tuning the carrier density at the interface. The dc current excitation was 1 $\mu$A. All magneto-transport measurements were performed with the magnetic field oriented perpendicular to the sample's plane.

**First-principles calculations:** The *ab initio* calculations have been performed using the plane wave code VASP [36, 37] and the generalized gradient approximation (GGA) [38] for the exchange-correlation energy. More details are given in the supplementary information S11.

# Aperiodic quantum oscillations in the two-dimensional electron gas at the LaAlO$_3$/SrTiO$_3$ interface

Km Rubi,[*] Michel Goiran, and Walter Escoffier
*LNCMI-EMFL, CNRS, INSA, UPS, 143 Avenue de Rangueil, 31400 Toulouse, France*

Julien Gosteau, Rémi Arras, Benedicte Warot-Fonrose, Raphaël Serra, and Etienne Snoeck
*CEMES, Université de Toulouse, CNRS, UPS, 29, rue Jeanne-Marvig, F-31055 Toulouse, France*

Kun Han, Shengwei Zeng, Zhen Huang, and Ariando[†]
*Department of Physics and NUSNNI-Nanocore, National University of Singapore, 117411 Singapore*

(Dated: February 26, 2019)

# Supplemental information

## (S1) STRUCTURAL CHARACTERIZATION OF LAO/STO SYSTEM AND SAMPLE DIMENSIONS

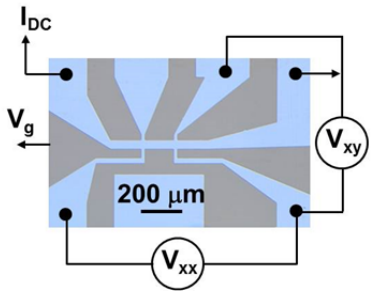

FIG. 1: An optical microscopy image of the LAO/STO device with Hall bar geometry.

In Scanning-TEM observations, the contrast is proportional to $Z^{1.8}$. Sr/La atoms exhibit brighter contrast than Ti/Al atoms due to their much higher Z number. The clear epitaxial growth of LAO on TiO$_2$-terminated STO bottom layer delineates that the LAO layer is the first deposited layer. The intensity profile show a progressive chemical transition from STO to LAO. It evidences a possible interfacial metallic (La,Sr)TiO$_3$ phase over 2 unit cell around the TiO$_2$-terminated STO layer. The parameter misfit between the substrate ($a_{STO} = 3.905$ Å) and the film ($a_{LAO} = 3.791$ Å) using a pseudo cubic description did not induced strain relaxations via interfacial misfit dislocations as no dislocation were observed along the entire TEM sample preparation. The structural distortions of the LAO (8 u.c.)/STO system was analysed by Geometric Phase Analysis (GPA). Results are presented on Fig.2. On the analysis presented, the STEM image has been acquired by scanning the electron beam with the fast direction from left to the right to minimise the scan distortion along the growth axis.

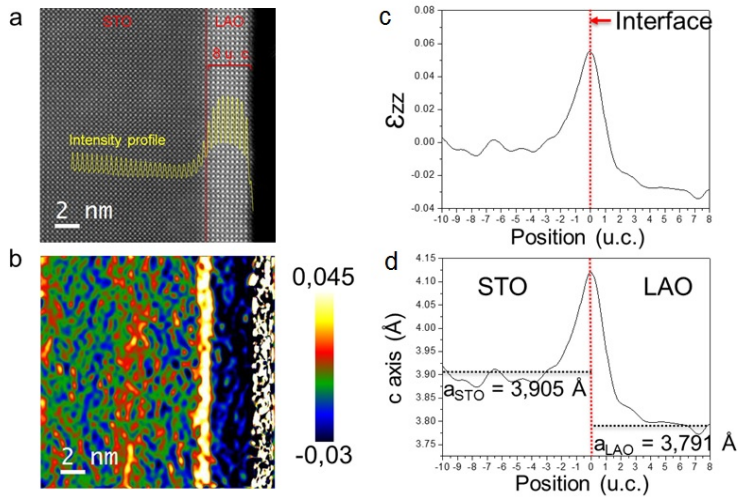

FIG. 2: Geometric phase analysis of the HAADF-STEM image. (a) HAADF-STEM image. The yellow curve corresponds to the intensity profile along the growth axis. (b) $\epsilon_{zz}$ strain map showing the strain characteristics along the c-axis, no in-plane strain ($\epsilon_{xx}$) has been observed. (c) Strain profile along the growth axis deduced from (b). (d) The c-axis variations and u.c.-volume changes in LAO (8 u.c.)/STO. Black dashed lines indicate the bulk parameters of STO ($a_{STO} = 3.905$ Å) and LAO ($a_{LAO} = 3.791$ Å). The Red dashed line indicates the LAO/STO interface.

Fig.2 shows $\epsilon_{zz}$ strain field along the growth axis deduced from Fig.2-a image via a GPA analysis. Fig.2-c presents the strain profile along the growth axis deduced from Fig.2-b. Fig.2-d presents the c-axis variations and u.c.-volume changes in the LAO (8u.c.)/STO system. The Red dashed line indicates the position of the TiO2-terminated STO layer. The geometric phase analysis on the LAO (8 u.c.)/STO system shows a c-elongation characteristic at the interface as seen on other investigations (SRXD in references [1–3]; TEM in reference [4]; STEM-HAADF in references [5–7]). This c-elongation is present in STO as well as in LAO with a value of 5.5% ($\sigma = 0.3\%$), in accordance with previous results [5, 6]. The width of the interface is between 3 and 4 u.c. With the mask used for the geometric phase analysis, the resolution of the strain map is 0.55 nm.

## (S2) BACKGROUND SUBSTRACTION

To check the influence of background substraction on $\Delta R_{xx}(B)$, we have used different polynomial expressions to fit the background. Although the amplitude of the oscillations may change, their 1/B-aperiodicity remains very robust. We observe the same 1/B values for the oscillations' extrema when the second derivative of $R_{xx}(B)$ $\left(-d^2 R_{xx}/dB^2\right)$ is plotted as a function of 1/B.

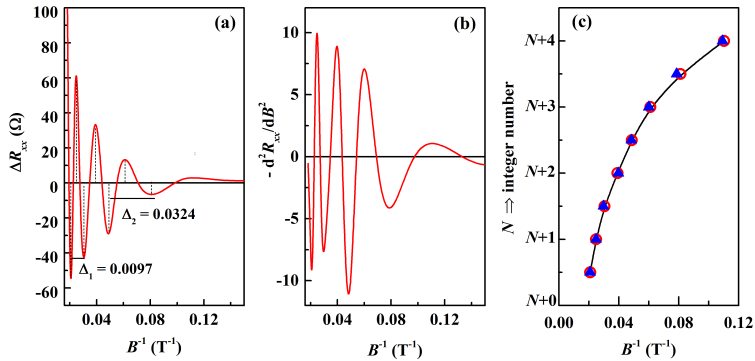

FIG. 3: (a) Inverse magnetic field dependence of the oscillatory part of $R_{xx}$ ($\Delta R_{xx}$) at T=1.6 K and $V_g = 0$ V after subtracting a monotonous polynomial background from $R_{xx}$. (b) The second order derivative of $R_{xx}(B)$ $\left(-d^2 R_{xx}/dB^2\right)$ at T=1.6 K and $V_g = 0$ V. (c) Comparison of Landau plots constructed from the extrema positions in $\Delta R_x x$ (red open symbol) and $-d^2 R_{xx}/dB^2$ (blue close symbol).

## (S3) CARRIER DENSITY ESTIMATED USING THE PLANE CAPACITOR MODEL

The change in the charge carrier density by applying a back gate voltage $V_g$ is calculated using the plane capacitor model:

$$\Delta n_{2D} = \frac{\epsilon_0 \times \epsilon_r(E) \times V_g}{e \times d}$$

where $d$ is the STO dielectric thickness, $V_g$ is the back-gate voltage, $e$ is the electron charge, $\epsilon_0$ is the vacuum dielectric permittivity and $\epsilon_r(E) \equiv \epsilon_r(V_g/d)$ is the relative dielectric permittivity of STO [8] (see inset of Fig.4).

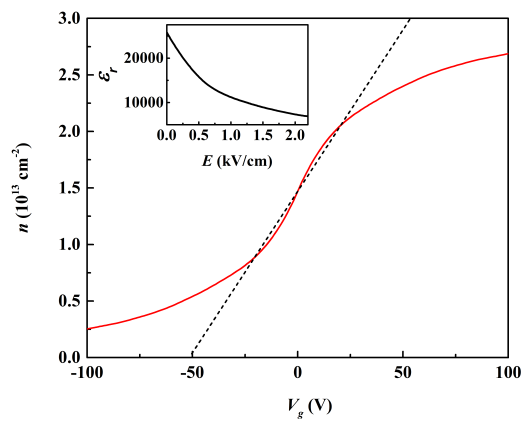

FIG. 4: Main panel: a comparison of $\Delta n_{2D}$ estimated using an electric field dependent STO dielectric permittivity (solid line) and a constant STO dielectric permittivity (dashed line) within the plane capacitor model. The inset shows the dielectric constant data as a function of electric field taken from Ref.[8].

## (S4) SIMULATION OF $\Delta R_{xx}(B)$ WHERE THE CARRIER DENSITY INCREASES WITH INCREASING MAGNETIC FIELD

$\Delta R_{xx}(B)$ is simulated using the simplified Lifshitz-Kosevich formula $\Delta R_{xx}(B) \propto A_T \times A_D \times \sin\left[2\pi\left(\frac{f(B)}{B} - 1/2\right)\right]$, where the frequency of the oscillations $f_{SdH}(B)$ is magnetic field dependent. Phenomenologically, the function $f(B)$ and the damping parameters $A_T$, $A_D$ are manually set in order to fit at best the experimental data. The carrier density $n_{2D}$ can be retrieved from the Onsager relation $n_{2D}(B) = \frac{g \cdot e}{h} f_{SdH}(B)$. We show the experimental ($V_g = 0$ V) and simulated data in Fig.5(a) and Fig.5(b) for initial carrier density $n_0 = 1 \times 10^{12}$ cm$^{-2}$ and $4.7 \times 10^{12}$ cm$^{-2}$, respectively. For $n_0 = 1 \times 10^{12}$ cm$^{-2}$, the experimental data can be fitted quantitatively with a monotonous increase of $f(B)$. However, the change in frequency with B is non-monotonous for $n_0 = 4.7 \times 10^{12}$ cm$^{-2}$.

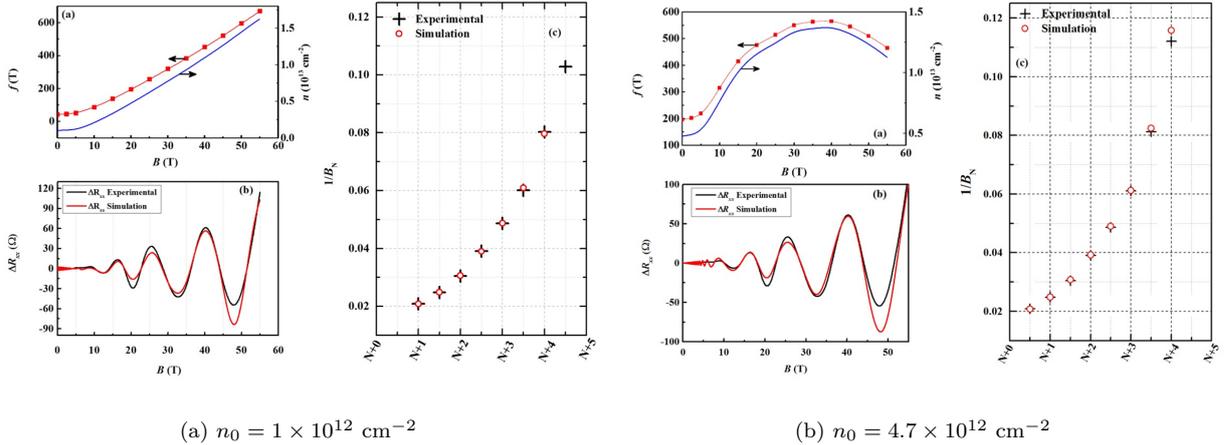

(a) $n_0 = 1 \times 10^{12}$ cm$^{-2}$      (b) $n_0 = 4.7 \times 10^{12}$ cm$^{-2}$

FIG. 5: Top-left panel: Oscillation's frequency (left y-axis) and carrier density (right y-axis) versus magnetic field. Bottom-left panel: Magnetic field dependent experimental $\Delta R_{xx}$ data for $V_g = 0$ V (black line) and simulated (red line) data. Right panel: Landau plots (1/B versus Landau level index) from experimental and simulated data.

While the phenomenological evolution of $f(B)$ fits the experimental data quantitatively, important issues must be highlighted: (i) the initial carrier density is much smaller than the Hall carrier density ($1.45 \times 10^{13}$ cm$^{-2}$), (ii) the simulations indicate a very large change of the carrier density (between 200% and 1000%) which is difficult to interpret. (iii) the large evolution of the carrier density will produce a non-linear Hall resistance in contrast with the experimentally observed linear behaviour.

We would like now to enrich the discussion by considering a two-fluid model, with the presence of high/low density and low/high mobility charge carriers. Within the two fluid model, the Hall resistance reads:

$$R_{xy}(B) = \frac{B}{e} \times \frac{(n_1 \cdot \mu_1^2 + n_2 \cdot \mu_2^2) + (n_1 + n_2) \times (\mu_1 \cdot \mu_2 \cdot B)^2}{(n_1 \cdot \mu_1 + n_2 \cdot \mu_2)^2 + (n_1 + n_2)^2 \times (\mu_1 \cdot \mu_2 \cdot B)^2}$$

Here, $n_1$ is replaced by $n_{2D}(B)$ as shown in fig.5(a) and 5(b). $n_2$, $\mu_1$ and $\mu_2$ are free parameters. The simulated Hall resistance is compared to the experimental one in figure6. The best fit for the experimental data is achieved using $n_2 = 1.45 \times 10^{13}$ cm$^{-2}$, $\mu_1 = 1000$ cm$^2$/V.s and $\mu_2 = 300$ cm$^2$/V.s. While the experimental data are well fitted in weak magnetic field regime ($0T < B < 10T$), a large deviation occurs for higher magnetic field.

To account for the experimentally observed 1/B-aperiodic SdHO, the model relying on a carrier density change as a function of magnetic field is therefore discarded.

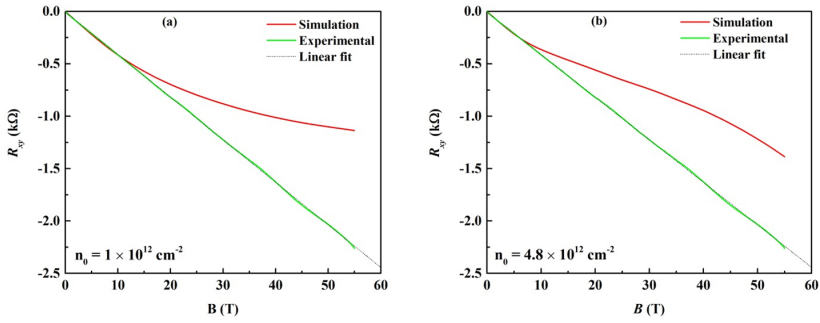

FIG. 6: Magnetic field dependent experimental and simulated Hall resistance data for (a) $n_0 = 1 \times 10^{12}$ cm$^{-2}$ and (b) $n_0 = 4.8 \times 10^{12}$ cm$^{-2}$. The dashed line shows a linear fit for experimental data.

### (S5) INCOHERENT 2D-NETWORK OF QUASI-1D CONDUCTION CHANNELS

In reference [9], G. Cheng *et. al.* suggested that aperiodic SdHO arise from the presence of an incoherent network of quasi-1D conduction channels, due to naturally formed domain walls at the 2D-LAO/STO interfaces. We evaluate here this hypothesis for our experimental data, following the work of C.W. Beenakker and H. van Houten [10]. In the ballistic conduction regime, the conductance is directly proportional to the number of electric sub-bands crossing the Fermi energy. This number is defined by both the electric confinement potential (i.e. which defines the lateral width of the conduction channel) and the magnetic field (which tends to increase and flatten the sub-bands energy towards the formation of Landau levels). The competition between the lateral and the magnetic confinement leads to aperiodic SdHO. Below, we consider two types of confining potentials: (i) a square-well potential of width $W$ and (ii) a harmonic potential with effective width $W_{eff}$. The effective mass of charge carriers in LAO/STO is assumed to be twice the bare electron mass, and the degeneracy is set to 2 (spin degeneracy only). The Landau plots and the fitting parameters are reported in Fig.7 and in table I, respectively.

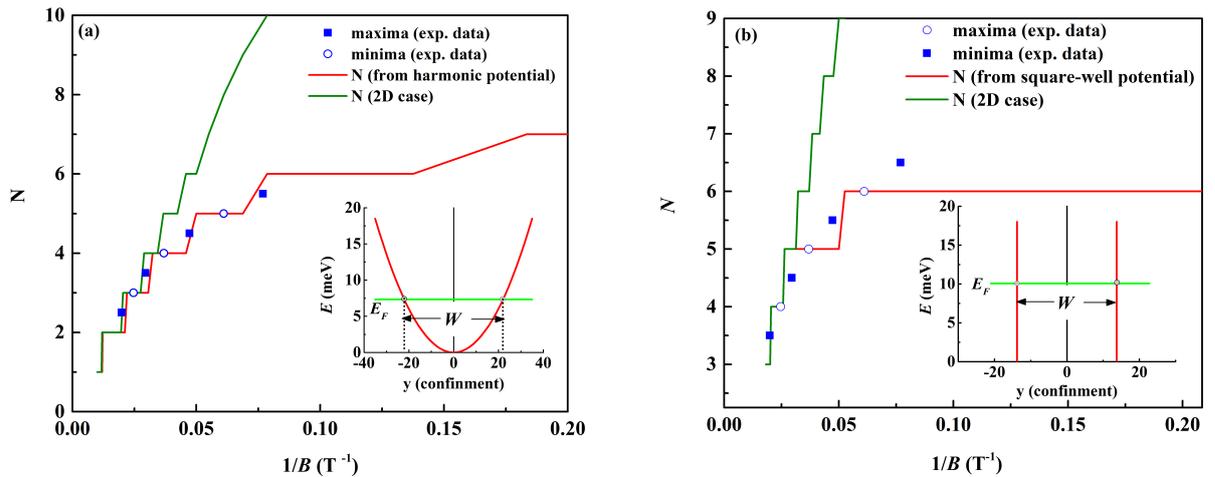

(a) Harmonic confining potential of effective width $W$

(b) Hard-wall confining potential of width $W$

FIG. 7: Landau plot: experimental data (blue symbols) and simulated data (red line for quasi-1D conduction channel and green line for 2DEG).

The Fermi energy, and resulting 2D carrier density $n_{2D} = \frac{g \cdot e}{h} f_{SdH}$, have been chosen to simulate at best the conduction regime cross-over, driven mainly by electric sub-bands at low magnetic field, and by Landau levels at high magnetic field. At high magnetic field, when the magnetic confinement dominates, the simulated Landau plot matches the 2D Onsager's relation as expected. Similar data fits can be obtained with slightly different fitting parameters, without altering the discussion below. The fitting parameter $N'$ is an arbitrary offset to the absolute Landau level index, since the quantum limit has not yet been experimentally reached in LAO/STO systems.

|              | $E_F$ (meV) | $n_{2D}$ ($\times 10^{12}$ cm$^{-2}$) | $W$ or $W_{eff}$ (nm) | $N'$ |
|--------------|-------------|---------------------------------------|------------------------|------|
| Square Well  | 10          | 8,4                                   | 27,5                   | 3    |
| Harmonic     | 7.3         | 6,1                                   | 44                     | 2    |

TABLE I: List of fitting parameters Fermi energy $E_F$, 2D carrier density $n_{2D}$, potential width $W_{eff}$ and arbitrary offset $N'$ for square well and harmonic potentials.

The fitting quality of the data relies on the profile of the confining potential (square well or harmonic potential): while the fit is qualitatively good for the harmonic potential, a square-well potential fails to produce a reasonable data adjustment. Without an independent knowledge of the input parameters (potential profile, $W$, $E_F$, $g$ and $m^*$), it is impossible to prove that the 1/B-aperiodic oscillations are linked to the presence of quasi-1D conduction channels. The model proposed by G. Cheng *et. al.* relies on the presence of an incoherent 2D-network of interconnected quasi-1D conduction channels naturally forming at the domain boundaries of SrTiO$_3$ close to the interface. The observation of clear aperiodic oscillations relies on the extreme condition that the channel width must be rigorously identical along the quasi-1D conduction channels. Any variations of the channel width translate into a fairly large change of the magneto-electric subbands spectra, which yields large blurring effect once averaged over the 2D interface. This situation seems difficult to be realized in practice, especially if one considers a finite thickness for the quasi-2DEG at the LAO/STO interface. In this case, one would rather deal with a 3D-network of interconnected quasi-1D conduction channels. In a 3D-space, the conduction channels are confined in two directions and their orientation can be arbitrary with respect to the interface plane. The crossings of the magneto-electric sub-bands with the Fermi energy are likely to be blurred, and therefore unobservable in the magneto-resistance.

## (S6) ROTH-GAO-NIU QUANTIZATION FIT

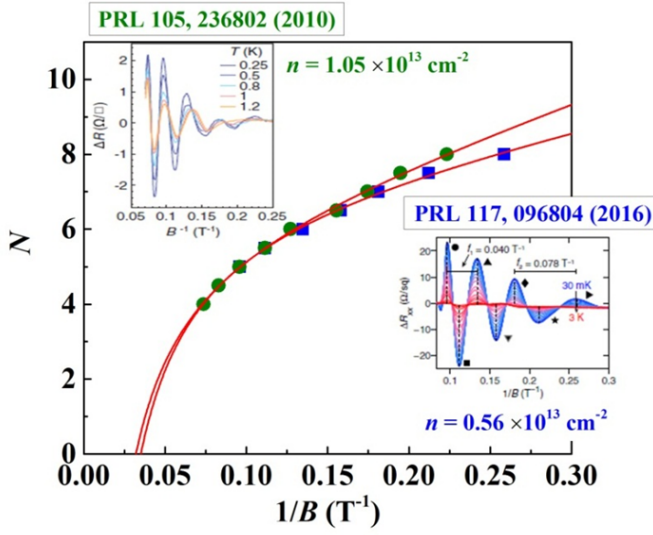

FIG. 8: Roth-Gao-Niu quantization fit to the data from literature Ref.[11] (green symbols) and Ref.[12] (blue symbols).

## (S7) GATE VOLTAGE DEPENDENCE OF ROTH-GAO-NIU FITTING PARAMETERS

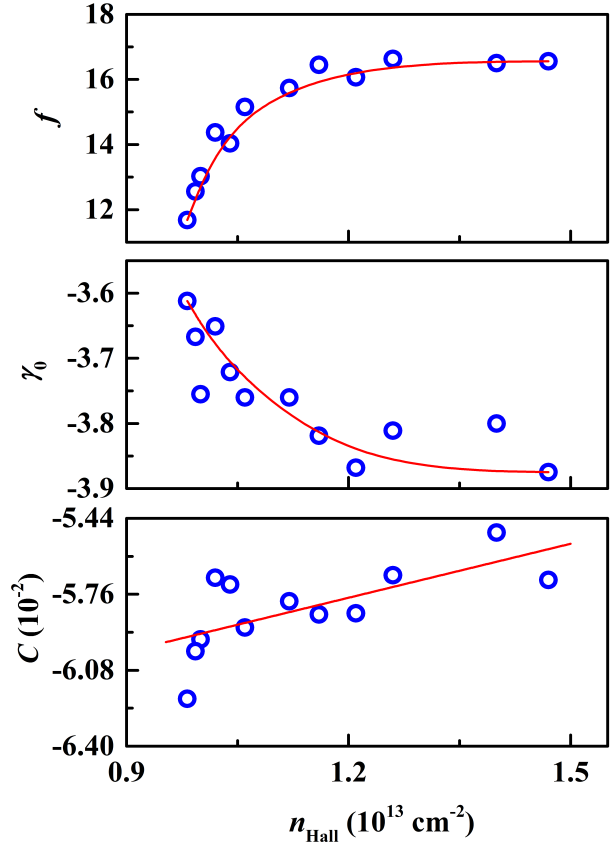

FIG. 9: The fitting parameters $f(E)$, $\gamma_0(E)$ and $C(E)$ as a function of Hall carrier density. Solid lines are guide for the eyes.

## (S8) FAST FOURIER TRANSFORMS (FFT) ANALYSIS

The fast Fourier transform (FFT) algorithm is applied to $\Delta R_{xx}$ versus $1/B$ data to evaluate the frequency spectra of the observed SdH oscillations. We show the FFT amplitude for few selected $V_g$ values (from $+20$ V to $-15.5$ V) in Fig.10. For $V_g = 0$ V, an intense peak $\beta$ corresponding to the frequency of $f = 55$ T is clearly noticed. In addition, another two peaks $\alpha$ and $\gamma$ are revealed at the left and right shoulders of the $\beta$ peak corresponding to the frequencies 22 T and 110 T, respectively. All the three peaks $\alpha$, $\beta$ and $\gamma$ are also observed for other $V_g$ values. While the peak positions remain almost constant when $V_g$ is varied from $+20$ V to $+5$ V, a substantial shift towards lower frequency is noticed for all the peaks as $V_g$ drops further (from $+5$ V to $-15.5$ V), in consistency with a global carrier density decrease.

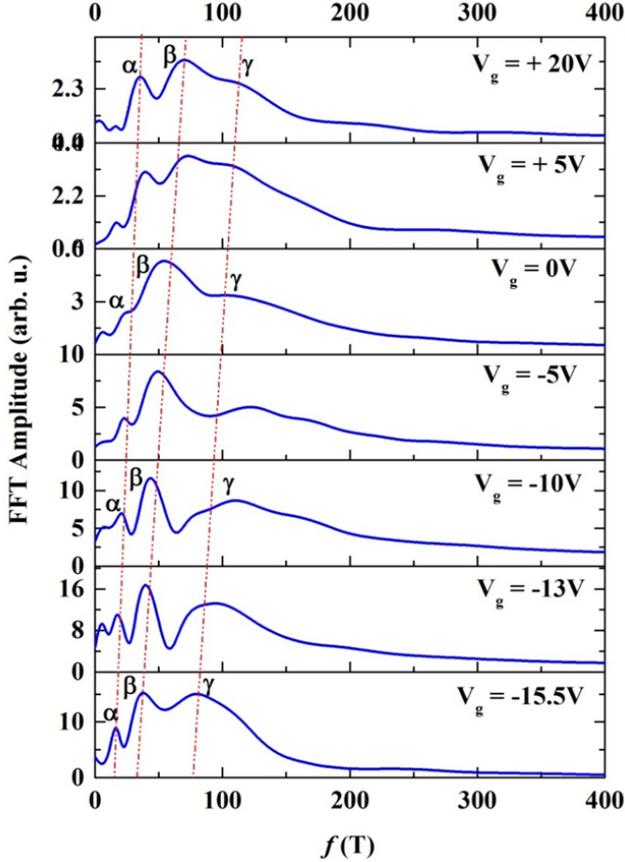
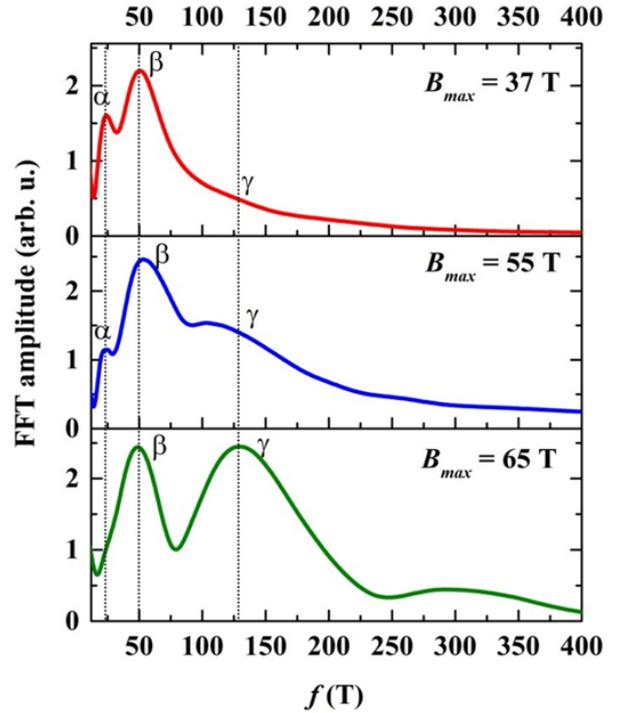

FIG. 10: Fast Fourier transform (FTT) amplitude of SdH oscillations for various back gate voltage values (from $+20$ V to $-15.5$ V). The dashed lines are the guide for the eyes and represent the shift in the FFT frequency with varying $V_g$.

FIG. 11: FTT amplitude of SdH oscillations for $V_g = 0$ V and different magnetic field ranges (a) $B_{max} = 37$T, (b) $B_{max} = 55$T and (c) $B_{max} = 65$ T.

Due to the large peak broadening, the interpretation of SdHO spectra is questionable. To check their reliability, we performed experiments for $V_g = 0$ $V$ within three different field ranges: (i) 37 T, (ii) 55 T and (iii) 65 T. The corresponding FFT spectra, shown in Fig.11, are all distinct from each other. While a $B_{max} = 55$ T data produces three peaks $\alpha$, $\beta$ and $\gamma$, only two peaks $\alpha$ and $\beta$ can be distinguished from $B_{max} = 37$ T data. Two distinct peaks $\beta$ and $\gamma$ are observed from the $B_{max} = 65$ T data and the $\gamma$ peak position is at higher frequency than that from the 55 T data. Considering the inconsistency in the spectra, it is clear that the FFT analysis fails to give reliable information of the oscillation's frequency. Indeed, besides the small number of periods, the monotonous increase of the oscillations' frequency at magnetic field provides a natural and mathematical explanation for the onset of the spurious $\gamma$ peak when the magnetic field range is extended to 65 T.

## (S9) HALL RESISTANCE DATA FITTED WITH TWO-FLUID MODEL

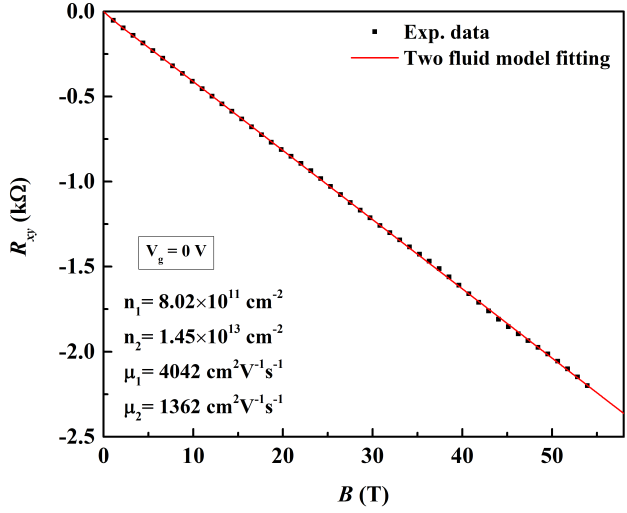

FIG. 12: Hall resistance $R_{xy}(B)$ with two-fluid model fitting. $n_1$ is kept fixed, $n2$, $\mu_1$ and $\mu_2$ are free fitting parameters.

## (S10) CYCLOTRON MASS ESTIMATION

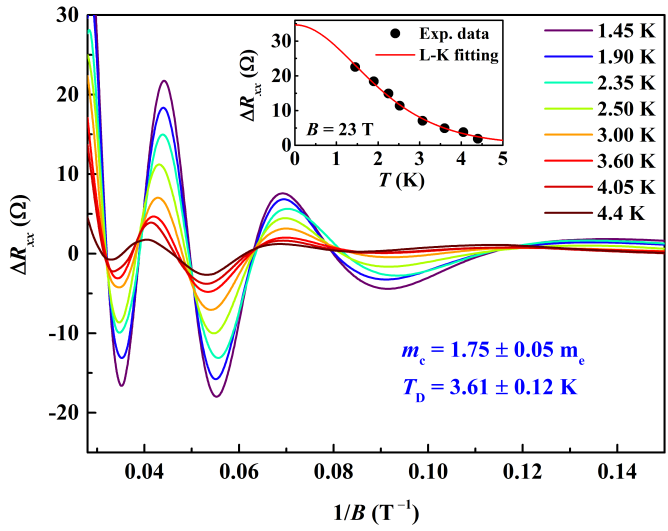

FIG. 13: Temperature dependent $\Delta R_{xx}$ (main panel) and Lifshitz-Kosevich equation fitting (inset) for oscillations amplitude at $B = 23$ T. The estimaed cyclotron mass is $m_c = 1.75 \pm 0.05\ m_e$.

## (S11) FIRST PRINCIPLE CALCULATIONS FOR ELECTRONIC BAND STRUCTURE

The DFT calculations have been performed using the plane wave code VASP [13, 14] under the Generalized Gradient Approximation (GGA) [15] to approximate the exchange and correlation energy. We have used the PAW [16] pseudopotentials with the following electronic configurations Sr (4s2 4p6 5s2), La(5s2, 5p6, 6s2, 5d1), Ti(3p6 4s1 3d3), Al(2s2 2p1), O(2s2 2p4), and a cut-off energy of 500 eV. We have used a symmetric slab with the following stacking layers: LAO(5 u.c.)/STO(20.5 u.c.)/LAO(5 u.c.)(001) (u.c. stands for unit cell). This heterostructure is composed by 2 identical perfect $TiO_2$/LaO-terminated interfaces and two perfect $AlO_2$-terminated surfaces. To avoid spurious interactions, the two surfaces have been separated by 15 Å of vacuum. The lateral lattice parameters in the directions [100] and [010] have been fixed to the experimental value measured for STO (3.905 Å) and the out-of-plane coordinates have been optimized. The LAO/STO structure has been relaxed until the atomic forces were below 0.03 eV/Å, using a $6 \times 6 \times 1$ Monkhorst-Pack [17] grid to sample the first Brillouin zone. The density of states (DOS) were calculated using a denser mesh, corresponding to a $12 \times 12 \times 1$ grid.

For a 5-u.c.-thick LAO layer, we found that 14 conduction bands of STO are crossing the Fermi level, and their total occupation corresponds to a charge density of $8.64 \times 10^{13}$ cm$^{-2}$, which spreads over 5 atomic layers of STO. Calculating higher charge density than the carrier density measured experimentally is common in the literature. This discrepancy may be partly explained by the fact that all the charges transferred at the interface may not be conductive. By adding an external electric field of 0.0024 $\mu$V/Å in calculation, the carriry density is reduced to $\sim 7.4 \times 10^{13}$ cm$^{-2}$, where the carriers start to populate the $d_{xz/yz}$ subbands. The calculations do not include the presence of defects which may affect the band population as well as the sub-band energies, and eventually introduces localization of the electrons. From the band structure, the $d_{xy}$ and $d_{xz}$ bands have the effective mass $m_{eff} \sim 0.5\ m_0$ along the $\Gamma - X$ direction, while the $d_{yz}$ has the effective mass $\sim 6.6\ m_0$. By considering the geometric mean, the averaged effective mass of these $d_{xz/yz}$ heavy bands is consequently calculated to 1.8 m$_0$. Our calculations are in agreement with those performed by Cancellieri, et al. [18].